\documentclass[aps,pre,twocolumn,showpacs,nofootinbib]{revtex4-2}
\usepackage{amsmath}
\usepackage{amssymb}
\usepackage{epsf}
\usepackage{graphicx}
\usepackage{color}
\begin{document}
\title{Force-Dependent Change of Structural Behavior Upon\\ Pulling a 
Grafted Flexible Polymer}
\author{J.~Matthew Smith}
\email[E-mail: ]{john.smith3@uga.edu}
\author{Michael Bachmann}
\email[E-mail: ]{bachmann@smsyslab.org}
\homepage[\\ Homepage: ]{http://www.smsyslab.org}
\affiliation{Soft Matter Systems Research Group, Center for
Simulational Physics, Department of Physics and Astronomy, The University of 
Georgia, Athens, GA 30602, USA}
\date{November 10, 2026}
\begin{abstract}
Manipulating polymer chains by means of experimental instruments such as 
optical tweezers has helped understand better the effects of molecular forces 
in macromolecular structures. In this study, we simulate the influence of a 
pulling force on structural phases of a coarse-grained model for flexible 
polymers using generalized-ensemble Monte Carlo methodologies. The recently 
introduced microcanonical inflection-point analysis method is employed to 
identify and classify structural phases of the system. Specific order 
parameters help characterize the transition behavior in response to the 
present pulling force. We show that, upon increasing the pulling strength, 
the coil-globule transition turns from second to first order, then merges 
with the liquid-solid transition. An intermediate phase, 
separating the coil and globule phases, is identified for certain force 
strengths. Clear evidence of the Gough-Joule effect is found as well. 
Transitions do not occur at all if 
the pulling force exceeds a certain threshold value. 
\end{abstract}
\maketitle 
\section{Introduction}
Experimental studies of polymer systems and computer simulations of 
models on various scales aid the better understanding of structure formation 
and structural transition 
behavior. In many cases, experiments and simulations complement each other. 
Whereas experiments offer deep insights into the real-world system, technical 
constraints often prevent systematic approaches that actually aid the 
understanding of generic system features that govern its behavior. Here is 
where computer simulations can help. Even though models are required that can 
only cover essential aspects of the actual problem, they enable a more 
systematic approach to studying these systems.

Optical tweezers~\cite{ashkin1,chu1} and other experimental techniques have 
been used with 
tremendous success to understand the impact of a mechanical pulling force upon 
linear polymers. These experiments make measurements of the forces possible 
that enable the formation of molecular structures in polymers such as proteins. 
This knowledge is essential, for example, for our understanding of protein 
folding processes and the cooperativity among parts of the system that allows 
it to transition from one macromolecular phase into another.

Previous work has been done to study the behavior of coarse-grained 
homopolymer models where the polymer is subject to an external force at one 
end~\cite{pincus1,halperin1,rosa1,polotsky1,polotsky2,%
grassberger1,jentzsch1,bell1}. These studies tended to focus on the behavior 
of the transition between the coil phase and the globule phase of the 
homopolymer. Literature on a comprehensive analysis of the phase 
structure in the state space of this problem is scarce. However, a more 
general phase portrait could already be obtained in extensive studies of a 
discrete polymer model~\cite{lsb14}. Among the major outcomes was the 
observation of the disappearance of the globular phase for increased tension 
and the absence of any transitions for forces above a certain threshold.

In this paper, we employ 
recently developed statistical analysis methodologies to systematically 
identify and classify the transitions between all 
phases. For this purpose, a generic coarse-grained polymer model of a 
flexible homopolymer grafted with 
one end to a substrate under the influence of a pulling force is used. The 
phase space is continuous and thus allows for the analysis of 
finite-size effects accompanying the major transitions. We also investigate 
the so-called tadpole phase in greater 
detail as it separates the random-coil and globular phases and only occurs 
under certain conditions. For our study, we performed replica-exchange Monte 
Carlo computer 
simulations~\cite{sw1,huku1,huku2,geyer1}. For the identification and 
classification of the structural 
transitions the polymer experiences when a pulling force of constant strength 
is 
present we employed the
microcanonical inflection-point analysis method~\cite{qb1}. Changes in phase 
behavior were studied systematically by 
repeating this process for various pulling strengths. 

We find that both the coil-globule and the freezing transitions shift to 
lower temperatures and eventually vanish as the applied force strength is 
increased beyond certain threshold values. Before it disappears, the 
coil-globule transition changes from second order to first order. The 
freezing transition is actually a transition sequence. As the temperature 
decreases, a solid core forms first inside the globular structures 
dominating the liquid phase. This solid-core formation process is a 
first-order transition. It is followed by a third-order surface restructuring 
process, in which the shell surrounding the frozen core crystallizes as 
well~\cite{qlkpwb1}. 
As an interesting by-product of our study we also observe the 
Gough-Joule 
effect~\cite{gough1,joule1,flory1}, in which the structure shrinks by 
increasing the 
temperature, even 
though a pulling force is applied. Another noteworthy result is the 
identification of a separate, intermediate phase between the coil and 
globular phases for certain temperature and force ranges. Referred 
to as tadpole structures, the key feature is the condensation of a
section of the coil into a compact substructure preceding the 
formation of a globule.

The paper is organized as follows: In Section~\ref{sec:modmeth}, we first 
introduce the coarse-grained model used in our study. We then briefly review 
the simulation methodology and describe the microcanonical and canonical 
statistical analysis methods used to identify, classify, and characterize the 
structural transitions. The results for an array of force values are 
discussed in Section~\ref{sec:res} and the paper is concluded by the summary 
in Section~\ref{sec:sum}. 
\section{Model and Methodology}
\label{sec:modmeth}
\subsection{Flexible Polymer Grafted to a Substrate 
Under the Influence of a Constant Pulling Force}
\label{subsec:mod}
\begin{figure}
\centerline{\includegraphics[width=8.8cm]{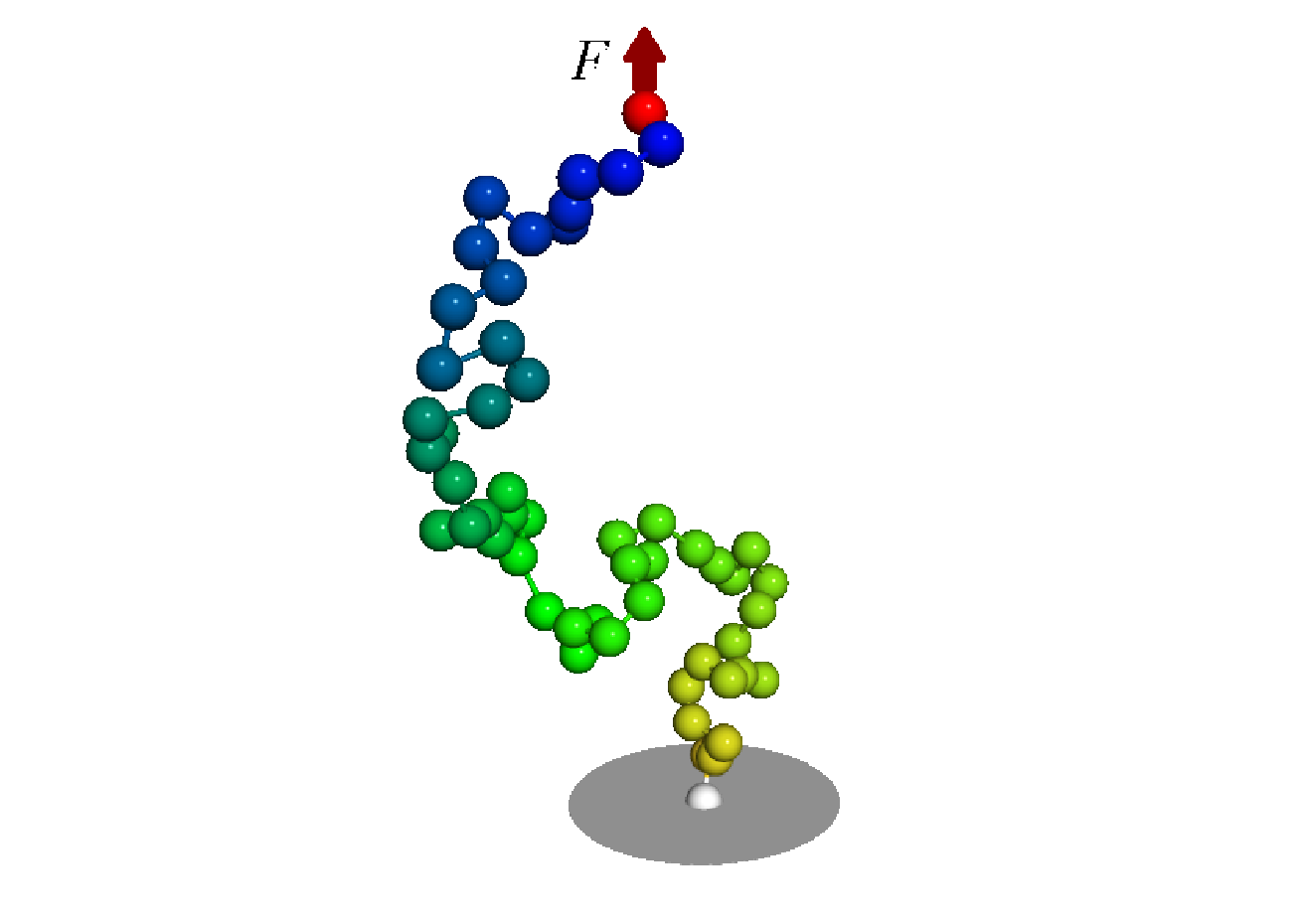}}
\caption{\label{fig:model}%
Example of a conformation of a polymer chain grafted to a planar, homogeneous 
substrate under the influence of a pulling force $F$ applied to the nongrafted 
end monomer (red). The other end monomer (white) is grafted to 
the substrate. The color gradient from the grafted to the pulled end is only a 
guide to the eye.}
\end{figure}

For our study we employ a generic coarse-grained model of a 
linear homopolymer. One end of the chain with $N$ monomers is grafted to a 
homogeneous substrate, which cannot be penetrated by the polymer but does not 
interact with the monomers in the polymer chain otherwise (see 
Fig.~\ref{fig:model}). We consider three 
principal contributions to the energy of any conformation 
$\mathbf{X}=(\mathbf{r}_1,\ldots,\mathbf{r}_N)$, where 
$\mathbf{r}_n=(x_n,y_n,z_n)$ 
(with $n=1,\ldots,N$) is the location of the $n$th monomer in the halfspace 
$z\geq 0$ above the planar substrate: Lennard-Jones interactions between all 
bonded and 
nonbonded pairs of monomers, the finite extensible non-linear elastic (FENE) 
potential representing the bonds between adjacent 
monomers~\cite{bird1,kremer1,milchev1}, and a linear potential associated with 
the constant pulling force $F$ on the non-grafted end of the polymer chain:
\begin{equation}
E(\mathbf{X})=E_\mathrm{LJ}(\mathbf{X})+E_\mathrm{FENE}(\mathbf{X})+
V_\mathrm{F}(z_N).
\end{equation}
The energy of the interactions that exist between all pairs of monomers is 
given 
by
\begin{equation}
E_\mathrm{LJ}(\mathbf{X})=
\sum\limits_{n=1}^{N-1}\sum\limits_{m=n+1}^N V_\mathrm{LJ}(r_{nm}),
\end{equation}
where
\begin{equation}
V_\mathrm{LJ}(r)=4\epsilon\left[ 
\left(\frac{\sigma}{r} \right)^{12} - 
\left(\frac{\sigma}{r}\right)^{6} \right]
\end{equation}
is the Lennard-Jones potential in standard form.
Here, $r$ is the distance between the monomers, $\sigma$ is the 
van der Waals radius, and 
$r_{0}=2^{1/6}\sigma$ is the location of the minimum of the 
Lennard-Jones potential. In our model we set $\epsilon=1$ to define our 
energy scale and chose $r_{0}=1$ to fix the length scale. We also introduce 
a cut-off at $r_\mathrm{c}=2.5\sigma$ for computational efficiency (hence 
$V_\mathrm{LJ}(r>r_\mathrm{c})=0$) and shift the potential by the 
constant value of $V_\mathrm{LJ}(r_\mathrm{c})$ to avoid a discontinuity at 
$r=r_\mathrm{c}$.

The FENE potential for all bonded monomers has the form
\begin{equation}
E_\mathrm{FENE}(\mathbf{X})=-\frac{1}{2}KR^{2}\sum\limits_{n=1}^{N-1}
\ln\left[1-\left(\frac{r_{n\,n+1}-r_0}{R}\right)^{2}\right]
\end{equation}
where $R$ gives the maximum displacement the monomers can have from the 
potential minimum, and $K$ fixes the energy scale of 
this interaction. These parameters are set to the standard values 
$R=3/7$ and $K=98/5$~\cite{qb2}.

To simulate the pulling force on the nongrafted end of the polymer, we 
introduce the simple linear potential
\begin{equation}
V_\mathrm{F}(z_N)=-F z_N,
\end{equation}
where $z_N$ is the perpendicular distance of the nongrafted end of the 
polymer from the planar substrate, and $F$ is the constant force applied to 
this end. 

The surface of the substrate 
is taken to be located in the plane $z=0$. The grafted end monomer resides 
at the 
coordinate origin $\mathbf{r}_1=\mathbf{0}$ and does not fluctuate, whereas 
the steric condition 
$z_n> 0$ is enforced on all other monomers to account for the presence of the 
impenetrable substrate.

In our case study, simulations for an array of different $F$ values were 
performed for a polymer chain with $N=55$ monomers. Various polymer models 
have been studied with this chain length before and serve as references. For 
example, the free flexible polymer has a unique, icosahedral ground state, 
which we can compare the crystalline structures in our study to. This chain 
length also represents a good compromise of computational efficiency, data 
quality, and statistically relevant size of the conformational phase space, 
which is essential for a detailed statistical analysis of stable structural 
phases.
\subsection{Parallel-Tempering Monte Carlo Simulations}
\label{subsec:mc}
For the simulation of our polymer system, we employed the replica-exchange 
(parallel tempering) Monte Carlo method~\cite{sw1,huku1,huku2,geyer1}. In this 
approach, simulation threads at $I$ different temperatures run in parallel. In 
each thread ordinary Metropolis Monte Carlo simulations are performed at fixed 
temperature, and after a certain number of Monte Carlo updates, replicas in 
neighboring threads are attempted to be exchanged. Independent simulations  
were 
performed for an array of constant force values in the interval 
$F\in [0.0,3.2]$. 
Depending on the given $F$ value, $I$ was chosen between $58$ and $61$ in the 
temperature range $T=0.04$ to $T=6.45$.

Monte Carlo updates of polymer conformations in each thread were proposed 
with a mixture of simple 
displacement moves and crank-shaft rotations. Displacement moves are 
performed by randomly selecting one of 
the (non-grafted) monomers, and then randomly generating a small displacement 
within a box of size $r_\mathrm{d}$ around this monomer. The value of 
$r_\mathrm{d}$ is adjusted in each thread in preliminary runs to maintain an 
acceptance rate for these updates between 20\% and 80\%.
The crank-shaft moves 
are performed by randomly selecting two monomers and rotating all monomers 
between them by a random angle about the axis connecting these two  
monomers. A Monte Carlo sweep consists of $N-1$ such updates since 
the grafted end monomer is bound in place. 

In Metropolis sampling, an update of the current conformation $\mathbf{X}$ is 
accepted according to the acceptance probability in temperature thread $i$ 
($i=1,\ldots,I$):
\begin{equation}
p_\mathrm{M}(\mathbf{X}\rightarrow\mathbf{X'})=\min\left(1,e^{- 
\left[E(\mathbf{X'}) - E(\mathbf{X})\right]/k_\mathrm{B}T_i}\right),
\end{equation}
where $\mathbf{X'}$ is the updated conformation. If the Monte Carlo 
move is rejected, the current conformation $\mathbf{X}$ is kept.

Every $2000$ sweeps, a swap of current conformations $\mathbf{X}_i$ and 
$\mathbf{X}_j$ between neighboring threads at temperatures $T_i$ and $T_j$ is 
attempted and accepted with the 
probability~\cite{sw1,huku1,huku2,geyer1} 
\begin{eqnarray}
&&p_\mathrm{PT}(\mathbf{X}_i\leftrightarrow\mathbf{X}_j;T_i,T_j)=\nonumber \\
&&\hspace*{1cm}\min\left(1,e^{ \left[E(\mathbf{X}_i) - E(\mathbf{X}_j) 
\right]\left(1/k_\mathrm{B}T_i - 
1/k_\mathrm{B}T_j \right)}\right).
\end{eqnarray}
The chosen temperature scheme ensures swapping rates between 
neighboring simulation threads are in the range 20\% to 80\%.

In each of our independent simulations at fixed force values, $2\times 
10^{9}$ sweeps per thread were generated, of which the initial 125 
million sweeps were 
used to equilibrate the Metropolis simulations in the individual threads. 
Canonical energetic and 
structural quantities were measured in the simulations and the energy 
histograms needed for the estimation of the density of states recorded. 
Numerical errors were estimated using the Jackknife binning 
method~\cite{miller1,efron1,mb1}. For each force value, the density of 
states, which is an essential ingredient of the generalized microcanonical 
inflection-point analysis method~\cite{qb1}, was determined by means of 
multi-histogram reweighting~\cite{hrew1,hrew2}.
\subsection{Microcanonical and Canonical Statistical Analyses}
\label{subsec:stat}
Generalized microcanonical inflection-point analysis is a method to 
systematically identify and classify phase transition signals in systems of  
any 
size~\cite{qb1}. It is based on the general assumption that transitions in a 
complex system are governed by its energetic and entropic properties. 
Least-sensitive inflection points in the microcanonical entropy
\begin{equation}
\label{eq:ent}
S(E)=k_\mathrm{B}\ln g(E)
\end{equation}
or its derivatives help identify transition points in energy space and 
determine their order.

The first derivative of the entropy is the microcanonical inverse temperature, 
$\beta(E)=dS(E)/dE$. Consequently, we denote subsequent derivatives of 
$S(E)$ by $\gamma(E)=d^{2}S(E)/dE^{2}$, 
$\delta(E)=d^{3}S(E)/dE^{3}$, etc. In this 
study, it turned out to be sufficient to analyze microcanonical quantities 
only up to the 
third derivative of the microcanonical entropy. 

In the generalized microcanonical inflection-point analysis method, 
transitions are broken into two main categories: independent transitions and a 
much rarer type of transition that only exists in finite systems known as 
dependent transitions. Independent transitions are the analog of traditional 
phase transitions. Dependent transitions can only occur in conjunction with an 
independent transition and can be considered a precursor for the independent 
transition it is associated with. In this study, dependent transitions were 
not found for the model parameter ranges considered.

The order of a transition is given by the order of the derivative of the
microcanonical entropy, in which the least-sensitive inflection point is 
found. 
An $n$th order transition is associated with a least-sensitive inflection  
point 
in $d^{(n-1)}S(E)/dE^{(n-1)}$. Consequently, a first-order transition 
is associated with an inflection point in the entropy, a second-order 
transition is characterized by a least-sensitive inflection point in the 
first derivative of the entropy, i.e., the microcanonical temperature 
$\beta(E)$, etc. 

From a technical perspective, it is convenient to identify 
least-sensitive inflection points of the microcanonical entropy and its 
derivatives by searching for an extremal point in the derivative one order 
higher than the derivative that possesses the inflection point indicating a 
transition.

Therefore, for an even-order independent transition of order $2n$ ($n$ a 
non-negative 
integer) with an inflection point at the energy value $E_\mathrm{tr}$, the 
$(2n)$th derivative of $S(E)$ possesses a negative maximum
\begin{equation}
\left.\frac{d^{2n}S(E)}{dE^{2n}}\right|_{E=E_\mathrm{tr}}<0
\label{eq:even}
\end{equation}
and 
for an odd-order independent transition of order $(2n+1)$, the $(2n+1)$th 
derivative of $S(E)$ possesses a positive minimum
\begin{equation}
\left. \frac{d^{2n+1}S(E)}{dE^{2n+1}}\right|_{E=E_\mathrm{tr}}>0.
\label{eq:odd}
\end{equation}
The density of states $g(E)$ was obtained by weighted 
histogram 
analysis~\cite{hrew1,hrew2}. In this scheme, 
\begin{equation}
\hat{g}(E)=\frac{\sum_{i=1}^{I}h_i(E)}{\sum_{i=1}^I M_i
\hat{Z}_i^{-1}e^{-\beta_{i}E}}
\end{equation}
serves as an estimator for the density of states and 
\begin{equation}
\hat{Z}_i=\sum_{E}\hat{g}(E)e^{-\beta_iE}
\end{equation}
is an estimator for the partition function at the simulation temperature 
$T_i$;
$M_i$ is the number of energy values calculated in the $i$th temperature 
thread of the parallel tempering simulation, and $h_i(E)$ is the number of 
energy values in the $i$th temperature thread that fall into the energy bin 
centered at energy $E$. 
These two expressions, when solved iteratively and initial guesses for 
$\hat{g}(E)$ and $\hat{Z}_i$ are chosen 
appropriately, converge 
towards the true values of the density of states and the partition function.

The estimate for the microcanonical entropy can be further 
improved by the B\'ezier method~\cite{mb1,bezier1,bezier2,gordon1}, which 
not only produces a smooth entropy curve, but it has the added 
advantage of allowing analytical derivatives to be taken.
For a set of $k$ ordered data points 
$(\mathbf{P}_0,\mathbf{P}_1,\dots,\mathbf{P}_k)$ the B\'ezier curve that 
smoothly connects these data points is given by
\begin{equation}
\mathbf{B}(t) = \sum_{i=0}^{k}\binom{k}{i}(1-t)^{k-i}t^{i}\mathbf{P}_i,
\end{equation}
where in our case $\mathbf{P}_i=(E_i,S_i)$ with $S_i=S(E_i)$ is the 
$i$th energy-entropy tuple obtained from the reweighting scheme. The curve 
parameter
$t\in [0,1]$ gives access to any point of the curve. The more convenient 
form~\cite{mb1,qlkpwb1}
\begin{equation}
S_\mathrm{bez}(E)= \sum_{i=0}^{k} \binom{k}{i} \!
\left(\frac{E_k-E}{E_k-E_0}\right)^{k-i}\!
\left(\frac{E-E_0}{E_k-E_0}\right)^{i} S_i.
\label{bezier_smoothing}
\end{equation}
provides a direct estimate for the B\'ezier-smoothed microcanonical entropy 
curve.
It is straightforward to calculate analytic derivatives of 
Eq.~(\ref{bezier_smoothing}) that can be easily implemented in an analysis 
script for the automated estimation of the extremal points conveniently used 
to identify the transition points [Eqs.~(\ref{eq:even}) and~(\ref{eq:odd})]. 
Statistical uncertainties in the estimation of 
the inflection points were taken care of by averaging over estimates obtained 
in independent simulation runs. The line thickness in the figures shown in the 
Results 
section was chosen to be larger than the actual standard deviation to 
accommodate these uncertainties.

For the canonical statistical analysis of structural transitions, we 
investigate thermal fluctuations of canonical order parameters at 
canonical inverse temperatures $\beta_\mathrm{can}$~\cite{rem1} near the  
transition points identified by microcanonical analysis. Following  
conventional 
statistical mechanics, the canonical average of a quantity $O$ is given by 
\begin{equation}
\langle 
O(\mathbf{X})\rangle(\beta_\mathrm{can})=\frac{1}{Z(\beta_\mathrm{can})}\int 
\mathcal{D}\mathbf{X}\, O(\mathbf{X}) e^{-\beta_\mathrm{can}E(\mathbf{X})},
\end{equation}
where $Z(\beta_\mathrm{can})=\int 
\mathcal{D}\mathbf{X} \exp[-\beta_\mathrm{can}E(\mathbf{X})]$ is the canonical 
partition function and $\mathcal{D}\mathbf{X}$ symbolizes the path-integral 
measure of the polymer conformations. Of more significant interest are the 
fluctuations of $\langle O\rangle$, though,
\begin{equation}
\frac{d\langle O(\mathbf{X})\rangle}{d\beta_\mathrm{can}}=
\langle O(\mathbf{X})\rangle\langle E(\mathbf{X})\rangle-
\langle O(\mathbf{X})E(\mathbf{X})\rangle,
\end{equation}
as extremal points of these curves may be associated with phase transitions in 
the finite system and, in combination with the transition points identified by 
microcanonical analysis, enable the structural characterization of the 
transitions.

Obvious choices of suitable order parameters are the square 
end-to-end distance 
$R_\mathrm{ee}^2(\mathbf{X})=(\mathbf{r}_N-\mathbf{r}_1)^2$ and the 
asphericity parameter
\begin{equation}
A(\mathbf{X})=\frac{R^2_{\mathrm{gyr},\perp}(\mathbf{X})}{R^2_{\mathrm{gyr},
\parallel}(\mathbf{X})}
\end{equation}
as the ratio of the contributions of the square radius gyration 
parallel and perpendicular to the surface of the substrate, 
respectively:
\begin{eqnarray}
R^2_{\mathrm{gyr},\parallel}(\mathbf{X})&=&\frac{1}{2N}\sum\limits_{n=1}^N 
\left[(x_n-x_\mathrm{com})^2+(y_n-y_\mathrm{com})^2\right],\nonumber\\
R^2_{\mathrm{gyr},\perp}(\mathbf{X})&=&\frac{1}{N}\sum\limits_{n=1}^N 
(z_n-z_\mathrm{com})^2,\nonumber
\end{eqnarray}
where $\mathbf{r}_\mathrm{com}(\mathbf{X})=(1/N)\sum_{n=1}^N \mathbf{r}_n$,
with components $x_\mathrm{com}$, $y_\mathrm{com}$, and $z_\mathrm{com}$, is 
the center of mass location of the polymer conformation $\mathbf{X}$.
\section{Results}
\label{sec:res}
In this section we summarize the results obtained by the different 
statistical 
approaches. First we employ the generalized microcanonical 
inflection-point analysis to identify and classify transitions and then 
perform 
structural analyses by means of order parameters to 
characterize the phases and transitions.
\subsection{Microcanonical Inflection-Point Analysis}
\label{ssec:micro}
\begin{figure}
\centerline{\includegraphics[width=8.8cm]{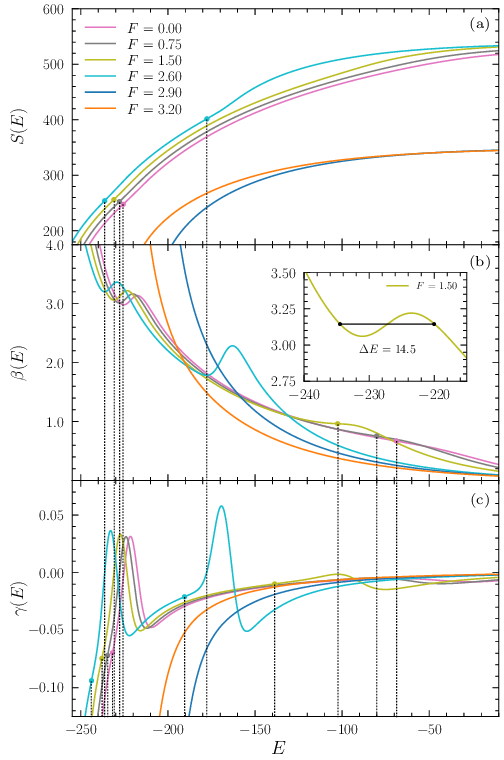}}
\caption{\label{fig:micropull}%
(a) Microcanonical entropy $S(E)$ and its derivatives (b) $\beta(E)$ and (c) 
$\gamma(E)$ as functions of energy for various values of the pulling force 
$F$. The inset in (b) shows an example of the horizontal Maxwell line 
and the definition of the latent heat $\Delta E$ for the first-order 
solid-core formation process at $F=1.5$.}
\end{figure}

The results for the microcanonical entropy and its derivatives obtained from 
the Monte Carlo simulations at different strengths of the pulling force 
are shown in Fig.~\ref{fig:micropull}. The dots mark the 
least-sensitive inflection points of the curves and are interpreted as 
transition signals according to the generalized microcanonical 
inflection-point analysis method~\cite{qb1}. All transition points and 
their classifications are listed in Table~\ref{tab:tr}. 

Even though a 
structural characterization of the phases separated by these transitions 
cannot yet be done at this stage, a few striking conclusions can already 
be drawn. The most obvious is the complete absence of transitions signals at 
the strongest pulling forces studied, $F=2.9$ and $F=3.2$, in the energy and 
temperature range we investigated. Entropic effects due to thermal activity 
are minimal and the polymer chain remains stretched out with very little room 
for fluctuations. Attractive monomer-monomer interactions cannot outweigh the 
stretching force either. Only the restraining force associated with the 
bonded potential competes with the pulling force and keeps the chain together.
\begin{table}[t]
\begin{tabular}{c|ccc}
\hspace*{5mm}$F$\hspace*{5mm} & \hspace*{5mm}$E_\mathrm{tr}$\hspace*{5mm} & 
\hspace*{5mm}$\beta_\mathrm{tr}$\hspace*{5mm} & 
\hspace*{5mm}order\hspace*{5mm} \\ \hline\hline
$0.00$ & $-231.82$ & $3.23$ & $3$ \\
       & $-225.31$ & $2.98$ & $1$ \\
       & $-68.79$  & $0.67$ & $2$ \\ \hline
$0.75$ & $-234.65$ & $3.29$ & $3$ \\
       & $-227.87$ & $3.012$& $1$ \\
       & $-80.19$  & $0.75$ & $2$ \\ \hline
$1.50$ & $-237.92$ & $3.36$ & $3$ \\
       & $-231.03$ & $3.06$ & $1$ \\
       & $-138.71$ & $1.17$ & $3$ \\
       & $-102.55$ & $0.96$ & $2$ \\ \hline
$2.60$ & $-243.91$ & $3.62$ & $3$ \\
       & $-236.33$ & $3.20$ & $1$ \\
       & $-190.49$ & $1.97$ & $3$ \\
       & $-177.63$ & $1.78$ & $1$ \\ \hline
$2.90$ & \multicolumn{3}{c}{no transition} \\ \hline
$3.20$ & \multicolumn{3}{c}{no transition} \\ \hline
\end{tabular}
\caption{\label{tab:tr} Table of transitions identified by microcanonical 
inflection-point analysis for different pulling 
forces.}
\end{table}
\begin{figure}
\centerline{\includegraphics[width=8.8cm]{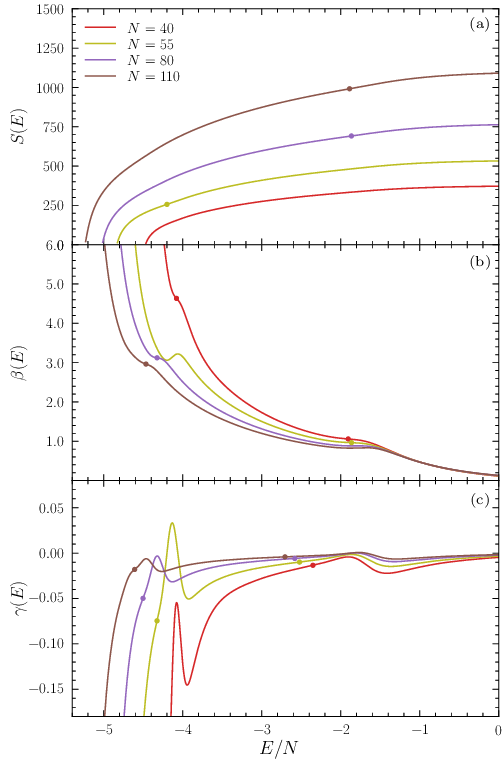}}
\caption{\label{fig:microlong}%
(a) Microcanonical entropy $S(E)$ and its derivatives (b) $\beta(E)$ 
and (c) 
$\gamma(E)$ as functions of the energy per monomer for different chain 
lengths at $F=1.5$.}
\end{figure}
\begin{figure*}
\centerline{\includegraphics[width=17.6cm]{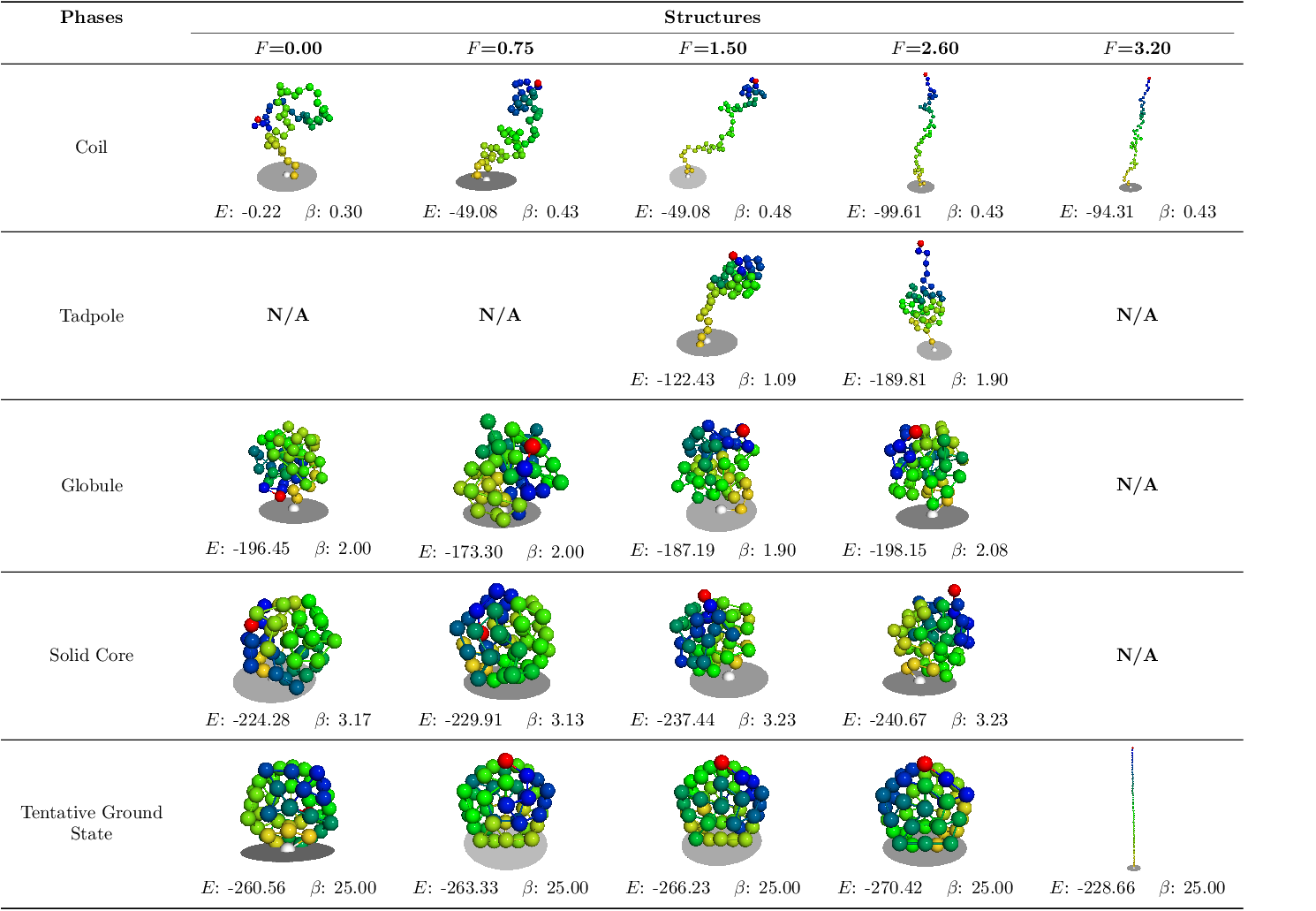}}
\caption{\label{fig:conftab}%
Representative conformations for the different structural phases.}
\end{figure*}

Conversely, for $F=0$, the system behaves like a free flexible polymer. The 
substrate only limits the available space for fluctuations in perpendicular 
direction, but the expected transitions are hardly affected. The polymer chain 
experiences the expected second-order coil-globule transition at 
$\beta\approx 0.67$ 
and the first-order freezing transition at $\beta\approx 2.98$. The latter is 
shadowed by a third-order transition in the solid phase at $\beta\approx 
3.23$. In fact, the formation of the icosahedral ground-state structure 
characteristic for this model is a two-step process~\cite{qlkpwb1}. Upon 
cooling out of the liquid phase, first a solid core is formed during the 
first-order transition, whereas monomers on the surface remain in a more 
liquid state, i.e., are more mobile and have not yet found their specific 
locations in the stable shell of the icosahedron surrounding the core. To 
complete this process, an additional transition is necessary. This 
surface-ordering transition is essentially a lower-dimensional liquid-solid 
transition, and it is of third order.

The latent heat of the first-order solid-core formation process can be 
obtained from the intersections of the Maxwell line and the original $\beta$ 
curve as shown for $F=1.5$ in the inset of Fig.~\ref{fig:micropull}. In this 
case $\Delta E\approx 14.5$. The latent-heat values of this transition at the 
other pulling forces we studied are similar. 

Therefore, the most interesting regime is the range of $F$ values for which 
the transition behavior is significantly influenced by the pulling force, but 
inter-monomer interactions and entropic effects are still sufficiently 
competitive. This can be readily seen by tracking the coil-globule transition 
upon increasing the pulling strength. For $F=0.75$ and $F=1.5$ it is still a 
second-order transition, albeit slightly shifted towards lower temperatures 
(higher $\beta$ values). However, for $F=2.6$ its character has dramatically 
changed. In this scenario it is a strong first-order transition. The 
transition point is now found near the liquid-solid transition. Following this 
trend, a merger of these transitions would be expected for stronger pulling 
forces, but, in fact, there is no trace left of the transitions for the 
other two cases studied, $F=2.9$ and $F=3.2$. 
It is likely that the merger occurs for force values between $F=2.6$ 
and $F=2.9$. Parallel-tempering simulations are known to fail in the presence 
of a strong first-order transition as is expected to occur in this regime. 
Further investigation is needed to conclusively investigate the behavior of 
the system in that region, but this would require a different simulational 
approach.

The first-order freezing transition is much less affected by pulling the 
chain, unless the pulling force exceeds a threshold value, at which the 
polymer cannot form sufficiently many 
monomer-monomer contacts needed for the creation of compact structures. For 
$F\le2.6$, liquid-solid transition points barely 
shift, nor do the third-order transitions in the solid phase, but for 
$F\ge2.9$, 
both transitions are gone.

Note that for $F=1.5$ and $F=2.6$, additional third-order signals occur 
between the coil and 
globular phases at slightly lower temperatures than  
the 
corresponding coil-globule transitions. Whereas these transitions are also 
consumed at stronger pulling forces, they represent an additional 
transition branch establishing an intermediate phase. Compared to the already 
discussed third-order transition in the solid phase, this additional signal 
in the liquid phase is much more subtle, but the close proximity to the 
liquid-globule transition suggests a similar mechanism: Monomers in clusters 
behave like a local liquid, whereas 
the other monomers respond to the external force and have detached from other 
monomers. This intermediate ``tadpole'' phase will be discussed in greater 
detail 
in the following structural analysis.

In order to verify the robustness of these results, we have also 
performed simulations for other chain lengths at $F=1.5$. These results are 
shown in Fig.~\ref{fig:microlong} and qualitatively confirm the discussion 
of the transition sequence for $N=55$, as expected. For the longer chains 
($N=80$, $N=110$), the coil-tadpole transition becomes first order, but the 
tadpole-globule transition remains of third order. The solid-core 
formation process, which is a first-order transition for $N=55$, is 
of second order for all other chain lengths studied. This is not particularly 
remarkable as $N=55$ is a ``magic'' number, i.e., our model favors the 
formation of icosahedral structures. This feature is not as significant in 
the other cases. The third-order surface restructuring transition is also 
present for the longer chains. It is of fourth order for $N=40$ as signaled by 
a corresponding inflection point in the next-higher derivative 
$\delta(E)=d^3S(E)/dE^3$ (not included in the figure).
\subsection{Structural Analysis}
\label{ssec:struc}
In this section, we discuss the structural properties of the phases separated 
by the transitions identified by microcanonical inflection-point analysis. 
We also briefly discuss the Gough-Joule effect~\cite{gough1,joule1,flory1} 
as an interesting 
side-effect also visible in our simulation data.
\subsubsection{Characterization of Conformational Phases}
Figure~\ref{fig:conftab} shows representative polymer structures for 
different pulling forces as found in the individual phases. For $F=0$, the 
ground-state 
conformation is icosahedral and reminiscent of the ground state for the free, 
nongrafted flexible polymer. Interestingly, it requires a substantial force 
($F>2.6$) to 
significantly deform the compact structure. In fact, it readily switches to 
a fully 
stretched-out conformation. Intermediate ground-state structures between 
these extreme conformations were not found for the force values simulated. 

In the solid-core phase the icosahedral core has formed, but the monomers in 
the shell still enjoy sufficiently entropic freedom to behave like a liquid. 
Consequently, rigid-body fluctuations allow the structures to tilt and dangle 
about the grafting point.
Globular conformations behave similarly, but do not have well-defined core. 
The effect from pulling is best visible in the coil phases, where the 
ordering 
effect from pulling and the entropic disordering from thermal fluctuations 
compete with each 
other.

\begin{figure}
\centerline{\includegraphics[width=8.8cm]{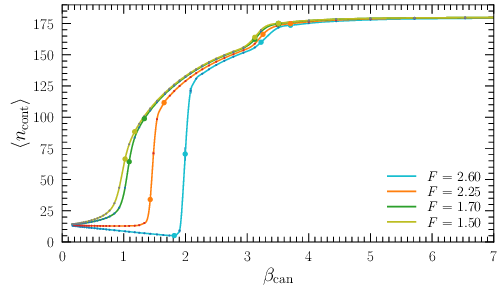}}
\caption{\label{fig:contactsBeta}%
Average of the number of monomer-monomer contacts 
$\langle n_\mathrm{cont} \rangle(\beta_\mathrm{can})$ as 
function
of the canonical inverse temperature $\beta_\mathrm{can}$ for four force 
values in the regime, where the 
tadpole phase exists. The dots mark the transition points 
identified by microcanonical inflection-point analysis.}
\end{figure}

Upon cooling out of the coil phase and if sufficient force is applied,
an 
additional phase exists between the coil and globule phases. The chain 
locally contracts as the formation of monomer-monomer contacts becomes 
favorable. We consider a monomer-monomer contact established, if the 
distance between two nonbonded monomers $n$ and $m$ is $r_{nm}<1.2$. 
The increase of the number of contacts as the temperature is decreased is 
obvious from 
Fig.~\ref{fig:contactsBeta}, where we can see that the number of contacts 
that 
must form to transition from the coil phase to the globule phase 
also increases 
with force. This is structurally confirmed by the decreasing square 
end-to-end distance shown in Fig.~\ref{fig:cansig}(b). 
The 
energetic difference from the constant force potential between the extended 
coil phase and the more compact globule phase becomes significant, requiring 
an increasing number of nonbonded monomer pairs to form contacts to dominate 
the constant force potential.

Polymer conformations in this regime are 
characterized by the formation of a local partial 
globule, with the 
remaining monomers forming into a coil, or coils if the globule is not at one 
end of the chain. If 
the chain was longer, multiple such local clusters would develop 
independently 
and form pearl-necklace type structures known from polyelectrolytes in poor 
solvent~\cite{holm1}.

The
coexistence of both coil-like and globular segments is 
characteristic of the so-called tadpole phase~\cite{pincus1,halperin1}. 
The two left-most dots in each curve shown in 
Fig.~\ref{fig:contactsBeta} 
correspond to the microcanonical inflection points associated 
with the coil-tadpole and the tadpole-globule transitions and are indicative 
of a significant change in the 
number of contacts 
as part of the chain collapse. In fact, the difference of $\langle 
n_\mathrm{cont}\rangle$ increases between the transition points from $F=1.5$ 
to $F=2.25$, further supporting the claim that the intermediate tadpole phase 
is necessary at these 
higher force values to build up the number of contacts required for 
the 
polymer to collapse into a globule under the force applied. For $F=2.6$, we 
already notice this difference to become smaller, suggesting  
the ultimate disappearance of this (and the other) phases for forces beyond 
the threshold force at which monomer-monomer contacts cannot form anymore.

Computer simulation studies of polymers in the constant end-to-end distance 
ensemble with variable force have also identified this 
phase~\cite{polotsky1,polotsky2,grassberger1,jentzsch1}. However, our results 
suggest that it is also present for unconstrained polymer chains in the 
canonical ensemble within a certain range of constant force values and 
temperatures.
\subsubsection{Gough-Joule Effect}
For the array of pulling forces we studied, Fig.~\ref{fig:cansig} shows the 
fluctuations of energy as well as the average square end-to-end distance as an 
appropriate order parameter for our system and its fluctuations, all as  
functions 
of the canonical inverse temperature. Taking a closer look at the inset of 
Fig.~\ref{fig:cansig}(b), we note that the end-to-end distance or the  
extension 
of the polymer chain shrinks upon heating towards $\beta_\mathrm{can}\to 0$,  
if 
the pulling force exceeds a certain threshold value (in our model $F\approx 
0.75$). For the free, but grafted chain ($F=0$), the end-to-end distance 
can increase up to the maximum possible length, limited only by the FENE  
bonds. 
This is the expected behavior in the random-coil phase, where thermal  
fluctuations 
experience entropic freedom in all three dimensions and the chain can explore 
the entire available structural state space. If a sufficiently strong pulling 
force is applied, though, the synthetically stretched chain can only 
fluctuate 
more freely in the directions parallel to the substrate. Now, in this 
situation, the FENE bonds actively pull back the monomers, and if 
$\beta_\mathrm{can}$ is sufficiently small, the overextended chain shrinks. 
This is known as the Gough-Joule effect~\cite{gough1,joule1,flory1}.
However, this is not a smooth process as a major phase transition interferes 
with it for a range of pulling force values, as will be 
discussed in the following section in more detail.
\subsubsection{Canonical Statistical Analysis of Conformational Phases}
\begin{figure}
\centerline{\includegraphics[width=8.8cm]{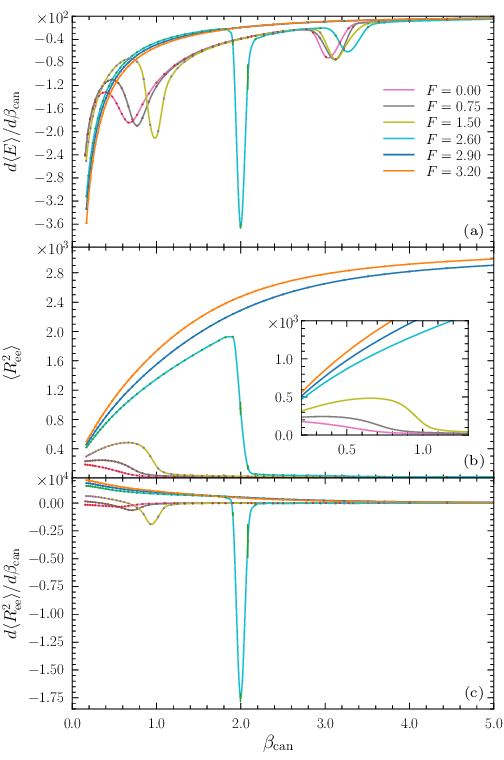}}
\caption{\label{fig:cansig}%
(a) Energy fluctuations $d\langle E\rangle/d\beta_\mathrm{can}$, (b) average 
square end-to-end distance $\langle R^2_\mathrm{ee}\rangle$, and (c) 
fluctuations of the end-to-end distance $d\langle 
R^2_\mathrm{ee}\rangle/d\beta_\mathrm{can}$ plotted as functions of the  
inverse 
canonical temperature for the pulling force strengths simulated.}
\end{figure}
\begin{figure}
\centerline{\includegraphics[width=8.8cm]{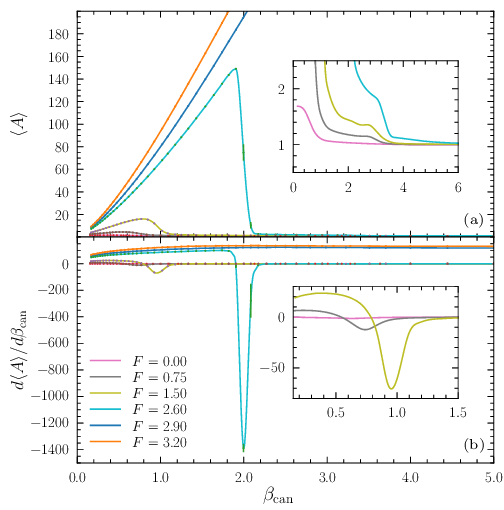}}
\caption{\label{fig:canasp}%
(a) Average asphericity parameter $\langle A\rangle$ and (b) fluctuations 
$d\langle A\rangle/d\beta_\mathrm{can}$ as functions of $\beta_\mathrm{can}$.}
\end{figure}

The canonical averages of square end-to-end distance $\langle 
R^2_\mathrm{ee}\rangle$ and asphericity parameter $\langle 
A\rangle$, as well as fluctuations of the 
energy and the structural quantities are shown in Figs.~\ref{fig:cansig} 
and~\ref{fig:canasp}.

Energy fluctuations measured in our simulations for the different pulling 
forces shown in Fig.~\ref{fig:cansig}(a) as functions of the inverse 
canonical temperature $\beta_\mathrm{can}$ clearly exhibit the influence of  
the 
pulling force on the major transitions known from the force-free flexible 
polymer ($F=0$), coil-globule and freezing. For reference, the corresponding 
minima in the energy fluctuation curve for $F=0$ are located at about 
$\beta_\mathrm{can}\approx 0.7$ (coil-globule transition) and at
$\beta_\mathrm{can}\approx 3.0$ (freezing) as known from former studies. The 
locations of these extrema coincide well with the transitions identified by 
microcanonical inflection-point analysis (see Table~\ref{tab:tr}), which 
suggests that the coil-globule transition is of second order, whereas freezing 
is a first-order transition, as expected. A weak pulling force $F=0.75$ leads 
to a shift of both transitions towards higher inverse temperatures. 

For force values $F>0.75$, we observe the Gough-Joule effect as discussed 
above, and transition signals keep shifting to higher $\beta$ values.
At $F=1.5$, the general phase structure still remains intact. 
However, the 
structural fluctuations at the coil-globule transition at $\beta\approx 0.96$ 
are significantly stronger compared to the results for smaller 
pulling forces. This is a first indication for a major change in the overall 
transition behavior. For $F=2.6$, both order parameters exhibit extreme 
fluctuations. As we know from the microcanonical analysis, the 
coil-globule transition is a first-order transition at this force. The 
canonical signal, which cannot distinguish between the coil-globule 
transitions and its third-order companion, is located near $\beta=2.0$. 
Compared to the $F=1.5$ case, the transition point has noticeably shifted 
toward a higher $\beta$ value, in fact by more than the shift of the freezing 
transition. For the latter, the fluctuation widths of the order parameters 
seem to become smaller. 

Even though the different shifts suggest a merger of the 
two transitions at some pulling force beyond $F=1.5$, we actually do not find 
any transitions in the fluctuation curves for the cases $F=2.9$ and $F=3.2$. 
The pulling forces are sufficiently strong to prevent monomer-monomer 
contacts to form and therefore suppress the formation of
phases governed by compact polymer structures.
\section{Summary}
\label{sec:sum}
We employed replica-exchange Monte Carlo simulations of a coarse-grained 
flexible polymer model to study the influence of an applied pulling force 
on one end of a polymer chain, whereas the other end was grafted to a 
substrate. Histogram reweighting enabled the estimation of the densities of 
states for an array of suitably chosen values of the pulling force. For 
each case, the microcanonical entropy, defined as the logarithm of the 
density of states, and its derivatives were used to identify and classify 
the structural transition the polymer experiences at the given applied 
force. This was achieved by means of the microcanonical inflection-point 
analysis method. The canonical structural analysis of the transition signals 
allowed us to characterize the structural phases. We also observed the 
Gough-Joule effect, where lateral entropic fluctuations can make the chain 
shrink by increasing the temperature despite the presence of a pulling force.

We find that below a certain threshold force both coil-globule and freezing 
transitions shift toward lower temperatures as the pulling force is 
increased. However, the magnitudes of the temperature shifts are different 
and the coil-globule 
transition point approaches the freezing point. The expected second-order 
transition for weak forces, the coil-globule transition, eventually turns into 
a first-order transition prior to its anticipated merger with the freezing 
transition. If the pulling force exceeds the threshold, both transitions 
completely vanish; the pulling strength suppresses the formation of  
monomer-monomer contacts necessary for transitions into compact structural 
phases. This confirms earlier results obtained in lattice polymer 
simulations~\cite{lsb14} that also showed the gradual disappearance of 
structural transitions as the pulling force strength is increased beyond a
threshold value. However, the Gough-Joule effect still persists 
under 
these conditions.

In addition to these major transition signals, we also identified third-order 
transitions, which can be attributed to certain stages of the structure 
formation processes that occur in the proximity of the major phase 
transitions. One occurs in what would usually already be called the solid 
phase. However, what is commonly called the liquid-solid transition in the 
thermodynamic limit, for a 
finite system it is better to be understood as a solid-core formation 
transition leaving the surface of the structure still in a liquid state. The 
surface layer then crystallizes out in a second step, which the 
microcanonical analysis identifies as a third-order signal. Similarly, the 
coil-globule transition is a two step process as well. Upon cooling, the 
second-order transition can, for a certain range of pulling strengths, lead 
from the coil phase to an intermediate 
ensemble of conformations, where individual globular and coil segments 
(``tadpoles'') coexist along the chain, 
similar to pearl-necklace conformations in other polymer systems. Only 
upon further cooling do these merge and the entire chain collapses by 
undergoing a third-order transition into the conventional globular phase.   

Microcanonical inflection-point analysis is a powerful tool capable of 
revealing entire transition sequences even for finite systems. It will help 
advancing this case study and may allow for the construction of the 
hyperphase diagram parametrized by the (inverse) temperature and the pulling 
force. Even though experimental verification of the character of 
structural transitions has been notoriously difficult, especially under 
equilibrium conditions, we hope that our results 
may encourage renewed interest in experimental studies of the structural 
behavior of polymers experiencing external forces under thermal conditions.
\begin{acknowledgments}
We thank the Georgia Advanced Computing Resource Center (GACRC) at the 
University of Georgia for providing computational resources.
\end{acknowledgments}
\vspace*{-3mm}
\section*{Data Availability}
The data that supports the findings of this article cannot be made publicly
available because the data is too large and no suitable public repository
exists for hosting in this field of study. The data are available upon
reasonable request from the authors.


\begin{thebibliography}{99}
\bibitem{rem1}
In our
simulations, we set the Boltzmann factor to $k_\mathrm{B}\equiv 1$. Therefore
inverse thermal energy $k_\mathrm{B}\beta_\mathrm{can}$ and inverse canonical
temperature $\beta_\mathrm{can}$ are measured in the same units throughout the
paper.
\end{thebibliography}
\end{document}